\documentclass[prl,aps,twocolumn]{revtex4}
\usepackage{graphicx}

\begin{document}

\title{Localization lengths of ultrathin disordered gold and silver nanowires}

\author{A. T. Costa Jr.}
\affiliation{Departamento de Ci\^encias Exatas, Universidade Federal de Lavras,
37200-000 Lavras, MG, Brazil}
\email{antc@stout.ufla.br}
\author{R. B. Muniz}
\affiliation{Instituto de F\'\i sica, Universidade Federal Fluminense,
24210-340 Niter\'oi, RJ, Brazil}
\email{bechara@if.uff.br}
\author{Jisang Hong}
\affiliation{Department of Physics and Astronomy,
University of California, Irvine, California, 92697 U.S.A.}
\author{R. Q. Wu}
\affiliation{Department of Physics and Astronomy,
University of California, Irvine, California, 92697 U.S.A.}

\begin{abstract}
  
  The localization lengths of ultrathin disordered Au and Ag nanowires are
  estimated by calculating the wire conductances as functions of wire lengths.
  We study Ag and Au monoatomic linear chains, and thicker Ag wires with very small cross
  sections. For the monoatomic chains we consider two types of disorder: bounded
  random fluctuations of the interatomic distances, and the presence of random
  substitutional impurities. The effect of impurity atoms on the nanowire
  conductance is much stronger. Our results show that electrical transport in
  ultrathin disordered wires may occur in the strong localization regime, and
  with relatively small amounts of disorder the localization lengths may be
  rather small.  The localization length dependence on wire thickness is
  investigated for Ag nanowires with different impurity concentrations.

\end{abstract}
\maketitle

Ultrathin metallic nanowires with decreasing diameters have been fabricated by
ingenious methods \cite{HDai}. Wires with average cross sections containing
just a few atoms have been reported recently \cite{Hong}. One extreme case is
the remarkable manipulation of single add-atoms on surfaces using a scanning
tunneling microscope, which allows wires to be built with atomic control of
its composition \cite{WHo}. These ultrathin nanowires are very important for
nanoelectronics. Not only they provide metallic contacts of nanoscopic
dimensions, but the electronic confinement in one dimension substantially
modifies the electrical characteristics, and offers exciting perspectives for
both basic and applied physics.

Most of the currently available techniques for manufacturing ultrathin
metallic nanowires produce structures with considerable amount of
inhomogeneities. Disorder effects in such systems strongly influences the
electrical transport, and may lead to localization. Under certain
circumstances, they can reduce the average conductance $\bar{g}$ very
effectively, making it fall off exponentially with the wire length $\ell$, for
sufficiently long wires. The localization length $\Lambda$ is determined by
the asymptotic decaying rate of $\bar{g}$ with $\ell$; more precisely,
$\Lambda^{-1} = -\bar{g}^{-1}\partial\bar{g} /\partial\ell$ for large values
of $\ell$. There are at least two additional characteristic lengths, relevant
for discussing electrical transport in ultrathin nanowires, the electronic
mean free path $\lambda$, and the average wire thickness or cross section
diameter $\bar{d}$. For relatively short nanowires with $\ell\ll\lambda$, the
transport is ballistic, whereas for sufficiently long wires with
$\ell\gg\Lambda$, it becomes strongly localized. Clearly, both $\lambda$ and
$\Lambda$ depend upon the amount and type of inhomogeneities present in the
system, and they also vary with $\bar{d}$.

Electrical transport through ordinary metallic structures are usually far from
the localization regime. On the other hand, non-interacting disordered
low-dimensional systems (with dimensionality $\leq$ 2) in the thermodynamic
limit have localized states only. They are, strictly speaking, insulators at
zero temperature \cite{Lee,Thouless}. Localization effects were observed in
very thin and short gold nanowires \cite{gold_nanow}, and a remarkably small
localization length ($\Lambda \leq$ 40 \AA) was estimated in such experiments.
Electrical conductance through nanocontacts between various metallic systems,
however, exhibits ballistic behavior \cite{onishi, tamura, garcia}.  Although
some degree of disorder is clearly present in these highly constricted
junctions, it is mainly effective on a relatively short length scale (of the
order of a few nanometers). Even when disorder is sufficiently strong for
localization to take place, the localization length needs to be compared with
the dimensions of the disordered system, especially when the latter are
relatively small. The observation of ballistic transport across those
nanocontacts indicates that either the disorder is rather weak, or the
localization length is much larger than the junction sizes.

The ability to fabricate thinner and longer disordered nanowires naturally
brings up the issue of localization. Here we address the following questions:
what is the order of magnitude of the localization length in disordered Ag and
Au nanowires? How does the localization length vary with the type and degree
of disorder that occur or may be introduced in these systems, and how does it
depend upon the wire average diameter? Such estimates are particularly
important when one considers the possibility of using these ultrathin metallic
wires in devices, and exploiting the quantum character of electronic transport
in such systems. In fact, disorder may be purposefully introduced to produce a
desired effect. It has been recently shown, for example, that electrical
current flowing through nanowires made of ferromagnetic disordered alloys can
become highly spin-polarized \cite{sf}. Behind this theoretical proposition,
however, lies the hypothesis that the spin diffusion length in such wires can
be much larger than $\Lambda$. Is this reasonable? In order to answer some of
these questions we have calculated the conductances of disordered Au and Ag
ultrathin wires, as functions of wire lengths. We start with monoatomic chains,
and consider two types of disorder: bounded random fluctuations of the
interatomic distances, and the presence of random substitutional impurities.
The conductance is calculated within linear response theory, with the use of
Kubo's formula. A multi-orbital tight-binding Hamiltonian with nine bands (the
five d-bands and the 4sp complex) is employed to describe the wire electronic
structure. The tight-binding parameters were obtained by least-square fitting
of first principles full potential linearized argument plane-wave (FLAPW) band
structure calculations of pure Au and Ag one dimensional wires. They reproduce
the lowest six FLAPW bands extremely well.

For an arbitrary wire it is useful to divide it in unit cells along the wire
axial direction, and express the conductance $\Gamma$ in terms of the
one-electron propagators connecting those cells. In units of the quantum of
conductance $\Gamma_0=\frac{2e^2}{h}$ we may write
\begin{equation}
\Gamma = {\rm Re}{\rm Tr}(\tilde{G}_{00}t_{01}\tilde{G}_{11}t_{10}-t_{01}
\tilde{G}_{10}t_{01}\tilde{G}_{10})\, , 
\label{conductance}
\end{equation}
where $\tilde{G}_{ij}$ represents the difference between retarded and advanced
one-electron propagator matrices, evaluated at the Fermi energy, connecting
cells $i$ and $j$; $0$ and $1$ labels two adjacent cells (the choice is
arbitrary due to current conservation), and $t_{01}$ is the tight-binding
hopping matrix between such cells. ReTr stands for the real part of the trace
over all orbitals and sites within an unit cell. For 1-atom thick chains, the
unit cell obviously contains a single atom.

We consider infinite wires with disorder restricted to a sector of length
$\ell$ along the wire axis. This is equivalent to assuming the current is
injected into and collected from the disordered nanowire by semi-infinite
perfect leads. The average wire conductance is calculated by taking into
account 4,000 random configurations of the disordered sector. The required
one-electron propagators are obtained as follows: first the tip Green
functions of the semi-infinite perfect leads are generated by numerical
methods that are, by now, well established and efficient
\cite{umersky,lsancho}. Then, the cells of the disordered wire sector are
placed, one by one, atop the injector lead.  This is carried out by using
Dyson's equation to switch on the hopping matrix between the added cell and
the substratum. Finally, the whole structure is connected, by the same
procedure, to the collector lead.

\begin{figure}
\rotatebox{-90}{\includegraphics[scale=0.3]{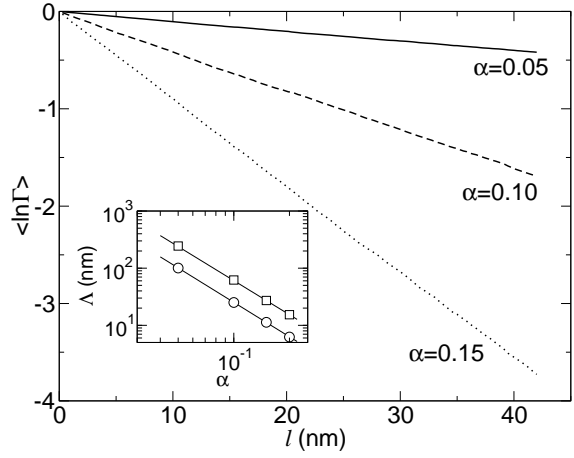}}
\caption{Average logarithm of $\Gamma$ as a function of wire length
 for different values of $\alpha$, obtained using Harrison prescription.
 The inset shows the localization length $\Lambda$ as a function of bond length 
 variation bounds $\alpha$, for Papaconstantopoulos (squares) and Harrison (circles) 
 prescriptions.}
\label{fig1}
\end{figure}

We shall start by examining disordered monoatomic chains in which the interatomic
distances $d$ vary randomly, within pre-established bounds, around the bulk
equilibrium value $d_0$. If the variations $\Delta d$ are not excessively
large, they may be modeled by scaling the hopping integrals.  Within certain
limits, the two center integrals $\gamma$ approximately follow power laws of
the type $\gamma(d)=(d/d_0)^n\gamma(d_0)$, where $n$ depends upon the pair of
orbitals involved. We have adopted two different sets of exponents: one
proposed by Harrison \cite{harrison}, and another by Papaconstantopoulos
\cite{papaconst}. More comprehensive scaling laws are presently available
\cite{papa2}, but those two prescriptions suffice for the kind of estimate we
wish to make. Figure \ref{fig1} depicts our results for the conductances of Au
and Ag disordered wires, calculated as functions of wire lengths for different
values of interatomic distance variation bounds $\alpha = d_m/d_0$; $d_m$
represents the maximum atomic separation.  The exponential decaying behavior
of $\Gamma$ with $\ell$, clearly shows that this type of disorder induces
localization.  The localization lengths $\Lambda$ deduced from the
corresponding decaying rates are plotted against $\alpha$ in the inset. Our
results show that $\Lambda$ is rather large for relatively low values of
$\alpha$. For instance, we found $\Lambda \approx 100$nm for $\alpha=0.05$
with the use of Harrison's scaling exponents.  Qualitatively, both
prescriptions give similar results: $\Lambda$ decreases with $\alpha$
approximately following a power law $\Lambda \approx \Lambda_0 \alpha^{-2}$,
with $\Lambda_0 \approx 0.62$nm and $\approx 0.25$nm, for the Harrison and 
Papaconstantopoulos scaling rules, respectively. Much larger bond length 
variations, corresponding to $\alpha\approx 0.3$, have been observed in 
stretched Au nanowires. However, there are strong indications that some of 
those apparently very large values
of $d$ are due to the presence of impurities \cite{Edison,Ugarte}.

Disorder caused by impurities leads to much stronger localization effects.  In
order to illustrate and investigate the role played by impurities in the
conductance characteristics of these nanowires, we consider substitutional Pt
atoms randomly distributed along the system. Pt in a noble metal matrix
represents a relatively weak scattering impurity potential. We simulate its
presence simply by modifying the atomic energy at the impurity site so as to
guarantee local charge neutrality. More elaborate ways of describing the
impurity potential are certainly possible, but we recall that our main
interest here is just to estimate the order of magnitude of the localization
length induced by it. Hence, such an approximate treatment of the impurity
suffices for our purposes.

\begin{figure}
\rotatebox{-90}{\includegraphics[scale=0.3]{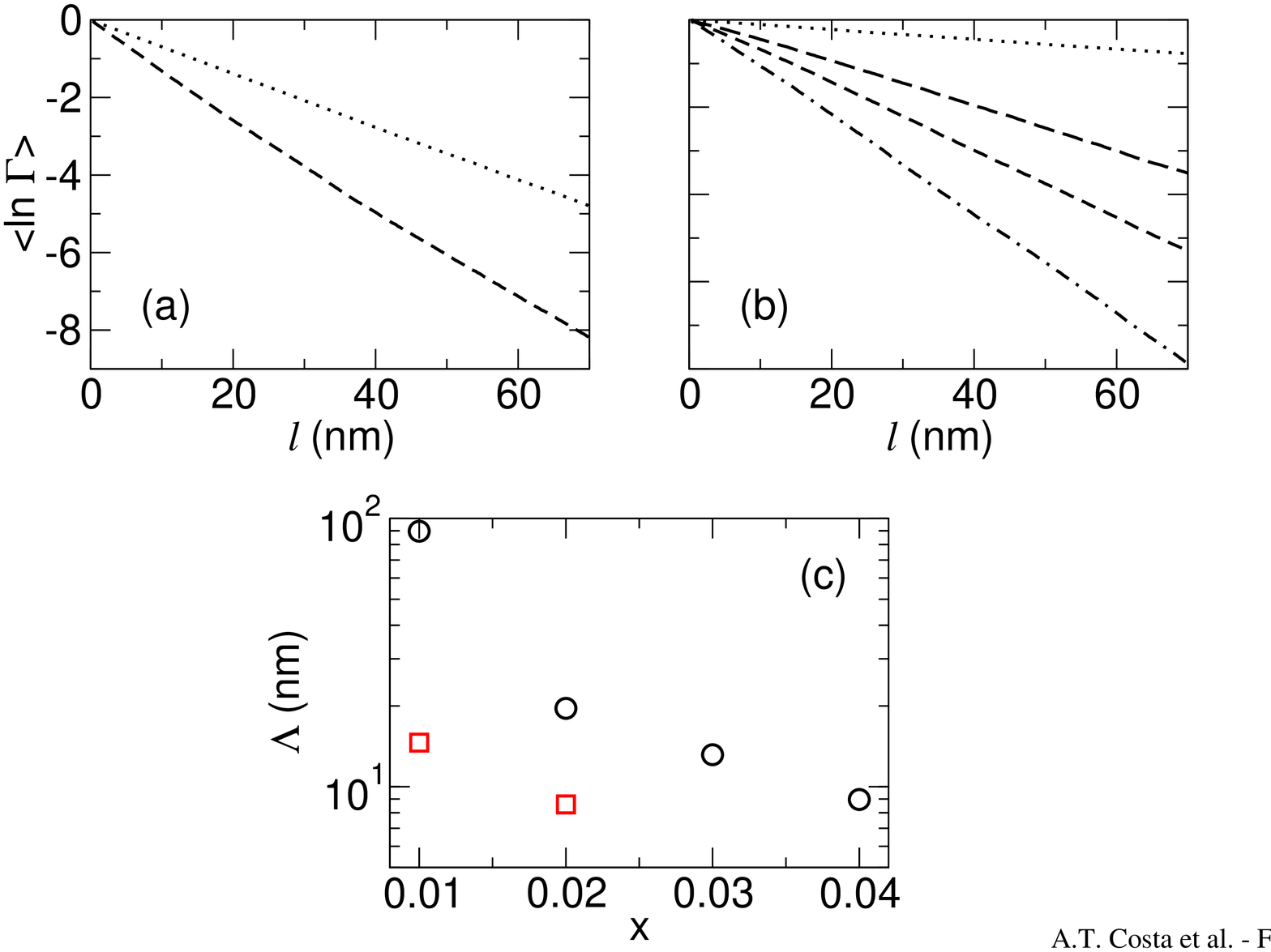}}
\caption{Conductance as a function of wire length for a Au (a) and Ag (b) atomic 
  chains with Pt impurities, for various impurity concentrations $x$. In (c)
  we show how $\Lambda$ varies with $x$ for the Au (squares) and Ag (circles) 
  wires. }
\label{fig_imp_chain}
\end{figure}

In figure \ref{fig_imp_chain} we show results for the conductances of monoatomic 
chains of Pt$_x$Au$_{1-x}$ and Pt$_x$Ag$_{1-x}$ disordered alloys, calculated
as functions of $\ell$, for different Pt concentrations x. We note in the
inset that $\Lambda$ decreases rapidly with x, as expected from the
one-dimensional character of the disordered wire. We found $\Lambda\approx
16$nm for just 1\% of Pt impurities in Au-based nanowires. It is also
noteworthy that $\Lambda$ becomes rather low for sufficiently large x,
reaching values of the same order of magnitude as that reported in ref.
\cite{gold_nanow}.  

\begin{figure}
\rotatebox{-90}{\includegraphics[scale=0.3]{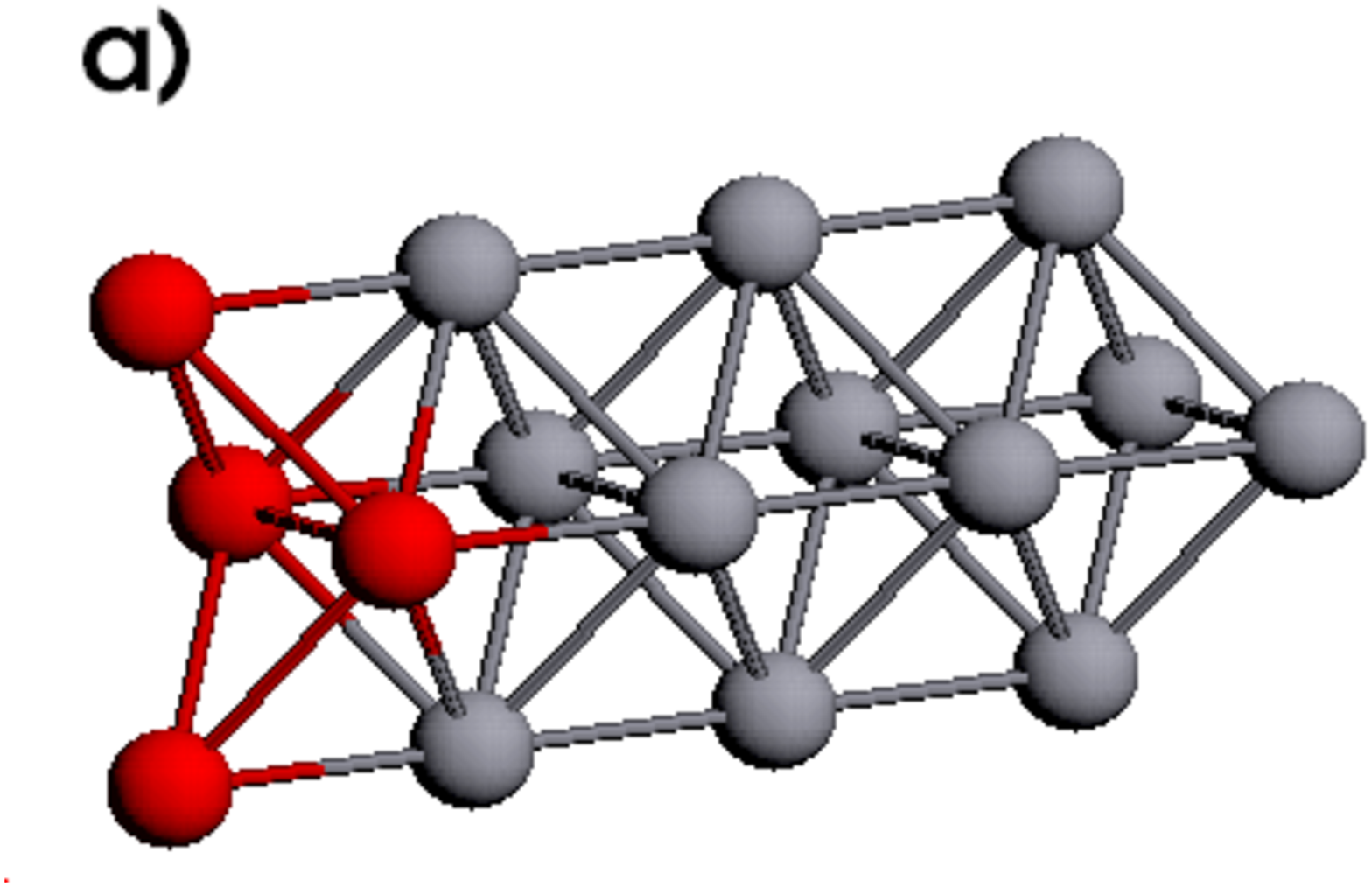}}
\rotatebox{-90}{\includegraphics[scale=0.3]{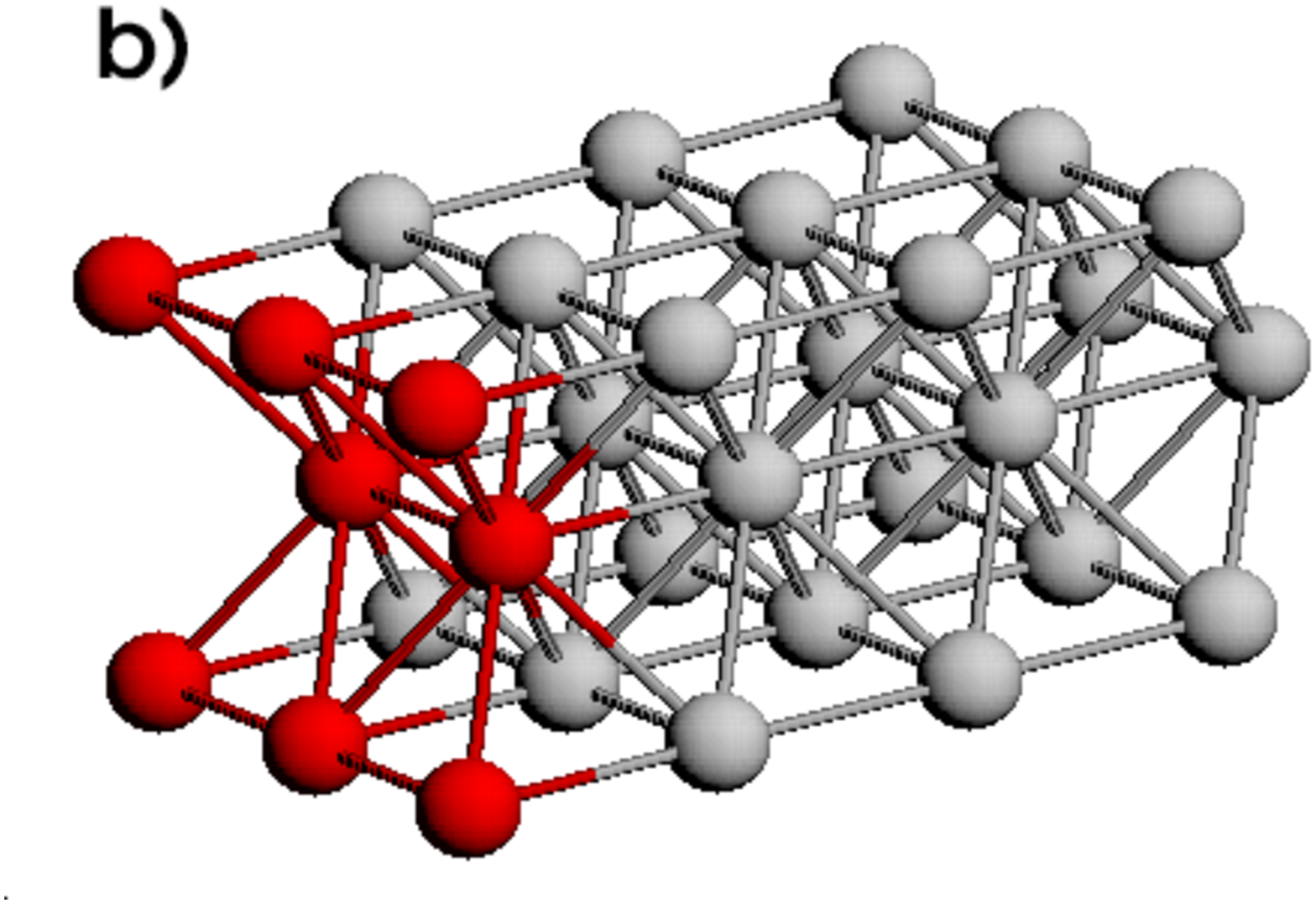}}
\caption{(Color online) Schematic representation of the 4 (a) and 8 (b) 
atoms cross-section wires. The atoms marked in red represent our choice 
of unit cell.}
\label{wires}
\end{figure}

One may rightfully argue that the effect of impurities is being overestimated
in those monoatomic linear chains, and that such systems are not so easy
to realize in practice. Moreover, it is rather difficult to observe
localization effects in systems with higher dimensionalities. Therefore, it is
important to investigate the dependence of the localization length on the wire
thickness \cite{Thouless}. Some kind of crossover from a localized to a
non-localized regime may occur as the wire cross section increases, for a
fixed impurity type and concentration. A particularly useful information is
the range of thicknesses in which a finite disordered wire is expected to be
found in the regime of strong localization. To gain some insight about this
matter, we have calculated the conductance of thicker Pt$_x$Ag$_{1-x}$
disordered wires as functions of $\ell$, for several values of x. We consider
wires with cross sections containing four and eight atoms, as schematically
shown in figures \ref{wires}a and b, respectively. The tight-binding
parameters for the 4-atoms-thick Ag wire were obtained by a crude fitting of
the band structure calculation reported in reference \onlinecite{Hong},
whereas for the 8-atoms-thick one we simply used bulk Ag parameters.
It is worth to mention that our results should not be much sensitive to
details of the electronic structure, since impurity-related effects are
expected to be much stronger than small variations in the hopping integrals.

\begin{figure}
\rotatebox{-90}{\includegraphics[scale=0.35]{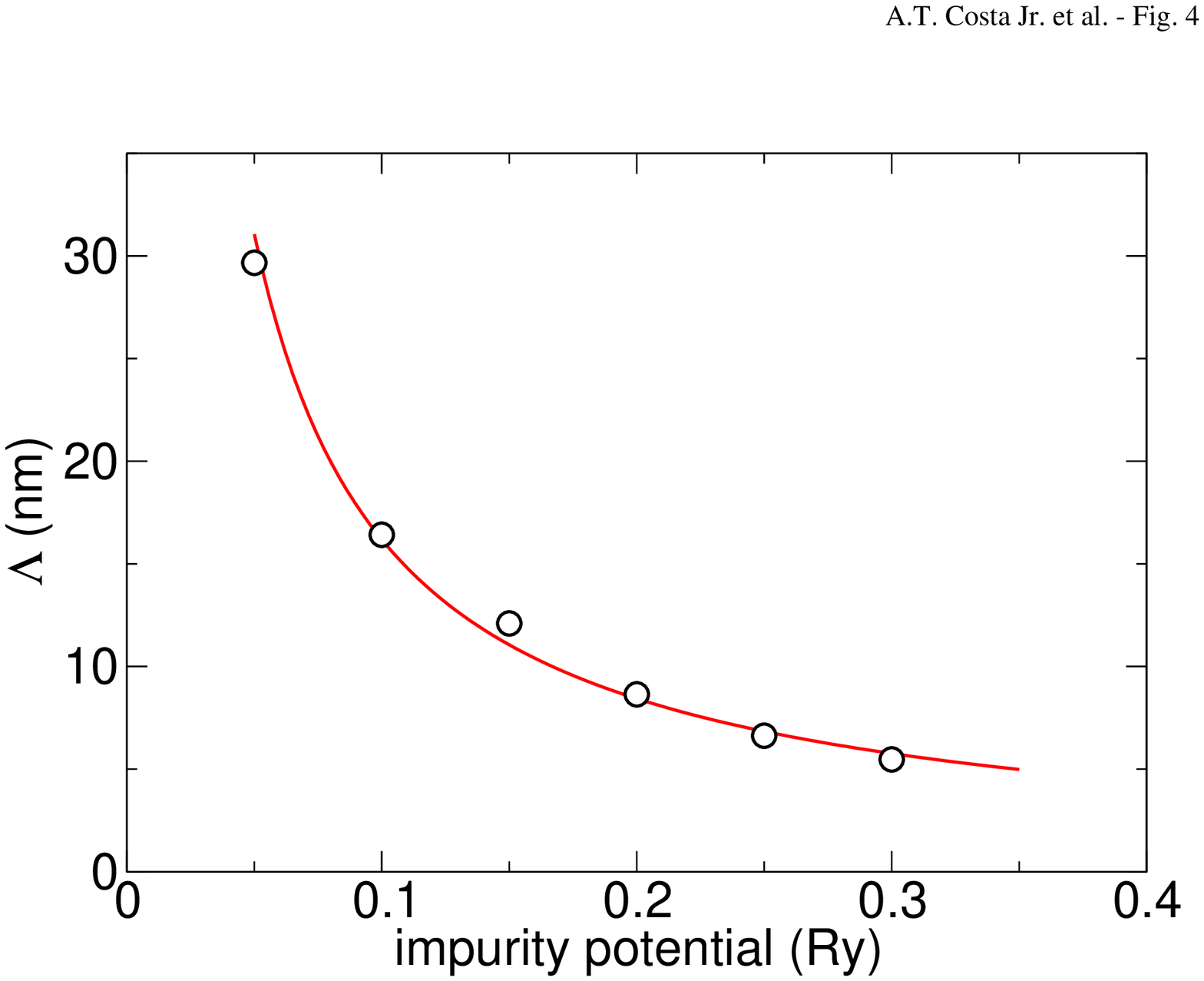}}
\caption{Localization length as a function of the impurity potential
for a 4 atoms thick Ag wire, with 2\% of randomly distributed impurities
(circles). The red curve is a fitting of the data to a power-law with
exponent $\approx 1$}
\label{cond_V_0}
\end{figure}

\begin{figure}
\rotatebox{-90}{\includegraphics[scale=0.3]{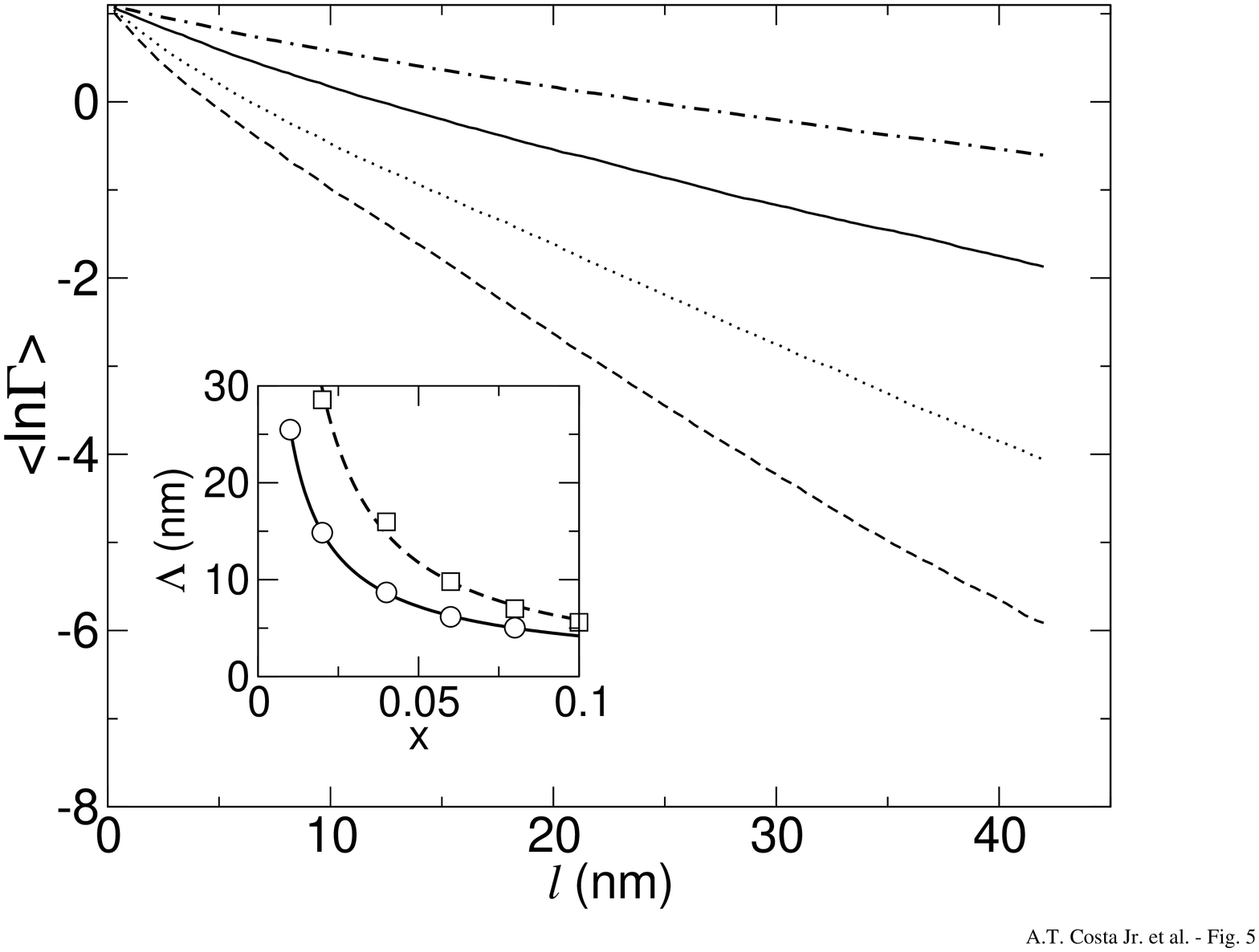}}
\caption{Conductance of a four atoms cross section Ag fcc(110) nanowire 
as a function of wire length for various impurity concentrations. The 
inset shows the localization length as a function of impurity concentration
for the four atoms (circles) and eight atoms cross section wires. The lines
are power-law fittings to the data (see text).}  
\label{cond_4}  
\end{figure}

Our results depicted in figure \ref{cond_4} show that the
localization lengths in these thicker (but still ultrathin) disordered wires
may be rather short for moderately low impurity concentrations. In both cases,
the localization lengths vary with the degree of disorder, apparently
following a power law. Even more intense localization effects are expected for
impurities which are associated with stronger scattering potentials as shown
in figure \ref{cond_V_0}. The most important information, however, is the fact
that those thicker wires also display localization on a nanometric length scale
for sufficiently large degree of disorder. 
These results support our
previous hypotesis that the localization length in ultrathin wires
made of ferromagnetic metallic alloys may be much smaller than the
spin diffusion length. With the increasing ability to control the
dimensions and composition of nanowires, we envisage the possibility
of using localization as an useful parameter to tailor electrical
characteristics in devices of nanoscopic dimensions. Disorered
nanowires form a rich laboratory for testing some very nice physical
ideas developed on extensive work following the classical article by
Abrahams et al. \cite{Gang_4}.

To summarize, we have calculated the conductance of disordered Ag and Au
nanowires using realistic band structures. Our results show that electrical
transport flowing through such systems may occur in a strong localization
regime. 
The localization lengths in monoatomic
linear chains can be remarkably small for moderately low impurity
concentrations. Localization persists in thicker disordered wires,
which also exhibit rather small localization lengths for modest
amounts of disorder. We hope our findings will stimulate further
investigation on these systems. If $\Lambda$ is comparable to the wire
length, to the electronic phase breaking or to the spin coherence
lengths, localization effects will be of primary importance in the
determination of the electronic behavior of such systems.

Enlightening discussions with D.L. Mills are gratefully acknowledged. 
A.T.C. and R.B.M acknowledge financial support from 
CNPq and the Milenium Institute for Nanoscience (Brazil). 
A.T.C also acknowledges financial support from FAPEMIG (Brazil).

\newpage

\end{document}